# On Thermal Radiation of Black Holes

A.A.Kyasov and G.V.Dedkov[1],

Nanoscale Physics Group, Kabardino –Balkarian State University, Nalchik

We calculate the photonic frequency–dependent absorption cross section and intensity of the Hawking radiation for a stationary non –rotating black hole using the electrodynamic membrane paradigm and Kirchhoff's law. A remarkable conclusion is that the use of geometrical optics limit for absorption cross-section and Kirchhof's law for the photonic luminosity of black hole provides a very good estimation of its total emission power.

1 INTRODUCTION

Following Hawking [1,2], a black hole in the vicinity of the event horizon will emit particles (photons, neutrinos, gravitons) as if it were a hot body with a temperature

$$T = \frac{\hbar c^3}{8\pi G M k_B}, \qquad (1)$$

where $\hbar$, $k_B$ and $G$ are the Planck's, Boltzmann's, and gravitational constants, $c$ is the speed of light in free space, and $M$ is the black hole mass. Particularly for a black hole without rotation and charge, the quantum emission rate $H(\omega)$ corresponding to a massless particle emitted in a wave mode with the frequency (energy) $\omega$, and the classical wave absorption cross-section $\sigma_a(\omega)$ are related by the Hawking's formula [2]

$$H(\omega) = \frac{\sigma_a(\omega)}{\exp(\hbar\omega/k_B T) \pm 1} \qquad (2)$$

Alternatively, the cross-section $\sigma_a(\omega)$ can be expressed in terms of the probability of particle birth in the vicinity of the event horizon.

According to (2) for example, the total emission power of photons is given by

$$I_{ph} = c\int_0^\infty H(\omega)(\hbar\omega^3/\pi^2 c^3)d\omega = c\int_0^\infty \frac{\sigma_a(\omega)}{\pi^2 c^3}\frac{\hbar\omega^3}{\exp(\hbar\omega/k_B T)-1}d\omega \qquad (3)$$

Eq. (3) is nothing but Kirchoff's law, whose applicability in the case of black holes was proven by Hawking.

In relation to Eq. (3), a very intriguing fact is that the geometrical optics approximation is not adequate for calculating $\sigma_a(\omega)$. Really, comparing the Schwarzschild radius $r_g = 2GM/c^2$ with the characteristic wave-length of thermal photons $\lambda_T = \frac{2\pi c \hbar}{k_B T}$, we obtain $r_g/\lambda_T = \frac{1}{8\pi^2} \ll 1$, that opposes the applicability condition of geometrical optics. Nevertheless, the latter has been currently used for estimating radiation intensity and life-time of black holes (see, for example,

---
[1] Corresponding author e-mail: gv_dedkov@mail.ru

Frolov [3], Shapiro and Teukolski [4]). This ambiguity is the first motivation point of our research.

Page [5] and Sanchez [6] have calculated the black hole absorption cross–section and luminosity corresponding to emission of photons and other particles using more accurate computational methods. However, despite the general agreement, there is a noticeable difference in the low–frequency asymptotics of the computed absorption cross–sections (see, for instance, Eqs.(19) in [5] and Eq.(2)in [6]. This difference may lead to differing evaporation times and other physical quantities relevant to black holes. Therefore, the obtaining long-wave asymptotics for absorption cross-section of a black hole seems to be of topical interest.

Another point of our study concerns electromagnetic properties of black holes. As has been realized by Novikov & Frolov [7], the event horizon of a black hole can be considered as a thin conducting sphere (membrane) with the surface resistance $R_H = \frac{4\pi}{c} = 377\,Ohm$. This simple result allows us to find the photonic absorption cross–section of black hole with respect to the low–frequency photons ($\lambda \gg r_g$) using the dipole approximation of classical electrodynamics. The corresponding photonic luminosity is then obtained applying the Kirchoff law. So, the aim of this work is to discuss the interrelation between the membrane paradigm and intensity of photonic partof the Hawking radiation on the one hand, and to compare different approximations for the photonic absorption cross-section, on the other hand.

## 2. THEORETICAL MODEL

According to the membrane paradigm [7], in the presence of external electromagnetic field the event horizon of a black hole behaves like a conducting surface. Introducing an effective thickness $\Delta r$ of membrane, and assuming the constant resistivity within the surface layer $\Delta r$, we can relate the total resistance $R_H$ of the black hole to the conductivity $\sigma$ of membrane using the model of two concentric spheres with radii $a_1 = r_g$, $a_2 = r_g + \Delta r$, where the spherical shell is filled by the homogeneous conducting matter (Batygin & Toptygin [8])

$$R_H = \frac{1}{4\pi\sigma}\left(\frac{1}{a_1} - \frac{1}{a_2}\right) \tag{4}$$

Assuming $\Delta r / r_g \ll 1$, we obtain

$$\sigma = \frac{\Delta r}{4\pi R_H r_g^2} = \frac{c}{(4\pi)^2}\frac{\Delta r}{r_g^2} \tag{5}$$

On the other hand, following Landau and Lifshitz [9], the dielectric permittivity of the corresponding conducting medium can be defined as

$$\varepsilon(\omega) = \varepsilon_0 + i\frac{4\pi\sigma}{\omega} \tag{6}$$

Eqs.(5), (6) make it possible to find the dipole polarizability of membrane as the polarizability of a spherical shell with the thickness $\Delta a = a_2 - a_1$, which is filled by the substance with dielectric permittivity $\varepsilon(\omega)$ (see Eq. (A2) from Appendix). The result is

3$$\alpha(\omega) = a_2^3 \frac{(1 - a_2^3/a_1^3)(2\varepsilon(\omega) + 1)(\varepsilon(\omega) - 1)}{2(\varepsilon(\omega) - 1)^2 - (a_2^3/a_1^3)(2\varepsilon(\omega) + 1)(\varepsilon(\omega) + 2)} \tag{7}$$

Substituting $a_1 = r_g$, $a_2 = r_g + \Delta r$ into (7) and expanding it up to a linear order in $x = \Delta r / r_g$, yields

$$\alpha(\omega) = x \frac{r_g^3}{3} \frac{(\varepsilon(\omega) - 1)(2\varepsilon(\omega) + 1)}{\varepsilon(\omega)} \tag{8}$$

Since $\lambda_T \gg r_g$, the corresponding absorption cross-section of electromagnetic radiation is given by the well-known formula [9]

$$\sigma_a(\omega) = \frac{4\pi\omega}{c} \alpha''(\omega) \tag{9}$$

where $\alpha''(\omega)$ denotes the imaginary part of $\alpha(\omega)$. Using (5), (8),(9) yields

$$\sigma_a(\omega) = \frac{r_g^2 x^2}{3}\left(2 + \frac{t^2}{\varepsilon_0^2 t^2 + x^2}\right), \quad t \equiv \omega/\omega_T = \frac{4\pi\omega}{c} r_g \tag{10}$$

Though Eq. (10) is obtained in dipole approximation, its important feature is that $\sigma_a(\omega)$ monotonically increases with increasing frequency and has a finite value at $\omega \to \infty$:

$$\sigma_a(\infty) = \frac{x^2 r_g^2}{3}\left(2 + \frac{1}{\varepsilon_0^2}\right) \tag{11}$$

Keeping in mind that at $\omega \to \infty$ the absorption cross-section asymptotically tends to (Landau and Lifshitz [10], Page [5], Sanchez [6])

$$\sigma_a(\infty) = \frac{27\pi}{4} r_g^2, \tag{12}$$

the value of $\varepsilon_0$ is naturally determined when equating (11) and (12)

$$\varepsilon_0 = \frac{2x}{\sqrt{81\pi - 8x^2}} \approx \frac{2x}{9\sqrt{\pi}} \tag{13}$$

Using (13) we can rewrite Eqs. (6) and (10) in the form

$$\varepsilon(t, x) = \frac{2x}{9\sqrt{\pi}} + i\frac{x}{t} \tag{14}$$

$$\sigma_a(t, x) = \frac{r_g^2}{3}\left(2x^2 + \frac{81\pi t^2}{4t^2 + 81\pi}\right) \tag{15}$$

Eq. (15) provides correct description both the long-wave and short-wave asymptotic of $\sigma_a(\omega)$. Eqs. (14),(15) can be useful in applications relevant to electrodynamics of black holes. So, inserting (15) in (3) yields the photonic luminosity of a black hole

$$I_{ph}(x, M) = \frac{1}{3072\pi^6} \left( \frac{\hbar c^6}{G^2 M^2} \right) f(x) \tag{16}$$

$$f(x) = \int_0^\infty \frac{t^3}{\exp(t) - 1} \left( 2x^2 + \frac{81\pi t^2}{4t^2 + 81\pi} \right) dt = 80.21 + 13x^2 \tag{17}$$

We see that $f(x)$ weakly depends on $x$ in the range $0 < x < 0.1$, relevant to the membrane paradigm. The corresponding evaporation time is then determined from $c^2 dM/dt = I_{ph}(x, M)$ and reads

$$\tau = \frac{1024\pi^6}{f(x)} \left( \frac{G^2 M^3}{\hbar c^4} \right) \approx 12270 \left( \frac{G^2 M^3}{\hbar c^4} \right) \tag{18}$$

Contrary to that, substituting in (3) the absorption cross–section (12), corresponding to the limit of geometrical optics, yields:

$$I_{GO} = \frac{27}{4} \pi r_g^2 c \int_0^\infty \frac{\hbar \omega^3}{\pi^2 c^3} \frac{1}{\exp(\hbar\omega/k_B T) - 1} d\omega = \frac{9}{20480\pi} \left( \frac{\hbar c^6}{G^2 M^2} \right) \tag{19}$$

$$\tau_{GO} \approx 2383 \left( \frac{G^2 M^3}{\hbar c^4} \right) \tag{20}$$

From (18) and (20) it follows $\tau/\tau_{GO} = 5.1$, but keeping in mind that photonic luminosity brings only about 20% in the total value of the Hawking radiation intensity [5], we can conclude that the use of estimations (19), (20) for the resulting black hole luminocity (involving all types of massless particles) and evaporation time is in quite fair agreement with our calculations.

3 COMPARISON WITH OTHER MODELS OF ABSORPTION CROSS–SECTIONS

It is expedient to compare (15) with the early known cross–sections calculated by Page [5] and Sanchez [6]. In our notations, the corresponding expressions for non–rotating black holes are given by

$$\sigma_a / r_g^2 = \begin{cases} \dfrac{t^2}{12\pi}, & t \ll 8\pi \\ \dfrac{27\pi}{4}, & t \to \infty \end{cases} \tag{21}$$

$$\sigma_a / r_g^2 = \begin{cases} \dfrac{27\pi}{4} - 4\sqrt{2}\pi \dfrac{\sin(\sqrt{27}t/4)}{t}, & t \geq 1.76 \\ 4\pi, & t = 0 \end{cases} \tag{22}$$

Contrary to (21), formula (15) predicts a finite value of the absorption cross–section at a zero frequency, and a larger numerical factor of the low–frequency expansion:

$$\sigma_a / r_g^2 = 2x^2/3 + t^2/3, \quad t \ll 1 \tag{23}$$

On the other hand, the membrane model predicts much lesser value for a zero-frequency cross-section (compare (22) and (23)). An overall view of the corresponding dependences is shown in Fig.1. Fig.1a compares (15) with (21) at $t < 0.1$, and Fig.1b compares (15) with (22) at $t > 1.76$. In spite of the difference in the long-wave asymptotics of the absorption cross-sections, the photonic radiation intensities which were computed using Eqs. (15) and (22) turn out to be close one another. Thus, integrating (22) according to (3) results in 27 to 36% difference as compared to (16). Therefore, as far as concerned the total radiation intensity, the membrane model is in a reasonable agreement with the calculation by Sanchez [6].

## 4 SUMMARY AND CONCLUSIONS

We have obtained a simple analytical formula for the photonic absorption cross –section of a black hole, using the electromagnetic membrane model of the event horizon. The resulting absorption cross –section in all frequency range is obtained using the calculated long –wave and the exactly known short –wave asymptotics. The additional parameters of the cross –section are introduced in a natural way. This allows us to determine the frequency –dependent dielectric permittivity of membrane, relating the static dielectric permittivity of membrane to its thickness and conductivity. The obtained expression for the dielectric permittivity can be useful for electrodynamics of black holes. Using Kirchhoff's law and the obtained absorption cross –section, we have calculated photonic luminosity and life –time of a black hole corresponding to the Hawking radiation. It follows that the membrane model predicts approximately five times larger photonic life –time of black hole as compared to that estimated in the limit of geometrical optics. Knowing that the total Hawking radiation intensity involving all types of emitted particles is approximately five times larger, the remarkable conclusion is that the use of geometrical optics limit for calculating absorption cross-section and Kirchhof's law for calculating photonic luminosity of black hole provides a very good estimation of its total emission power. This surprising result was not evident from the early beginning.

## APPENDIX

The static polarizability of a twin-layer dielectric sphere (Fig. 2) can be found solving the Laplace equation for the electric potential. If a dielectric sphere is placed in the homogeneous external electric field **E** being directed along $z$-axis (Fig.2), the Laplace equation $\Delta\phi = 0$ should be solved subjected to the boundary conditions of continuity $\phi$ and the normal projections of the electric displacement **D** at $r = a_1$ and $r = a_2$. In the outer space $r > a_2$ we finally obtain

$$\phi(r,\theta) = -Er\cos\theta + \frac{\alpha E}{r^2}\cos\theta \tag{A1}$$

$$\alpha = a_2^3 \frac{(1+2\varepsilon_2)(\varepsilon_1 - \varepsilon_2)a_1^3 + (\varepsilon_1 + 2\varepsilon_2)(\varepsilon_2 - 1)a_2^3}{2(\varepsilon_2 - 1)(\varepsilon_1 - \varepsilon_2)a_1^2 + (\varepsilon_1 + 2\varepsilon_2)(\varepsilon_2 + 2)a_2^3} \tag{A2}$$

where $\alpha$ is the static polarizability. Using the transformations $\varepsilon_i \to \varepsilon_i(\omega)$, $i = 1,2$ in (A2) we immediately obtain the dynamic polarizability $\alpha(\omega)$. The corresponding long –wave approximation is valid since $\omega a_2/c \ll 1$. Eq. (7) follows from (A2) at $\varepsilon_1 = 1, \varepsilon_2 \to \varepsilon(\omega)$

To obtain the polarizability of a thin dielectric spherical shell with thickness $\Delta a$, one must set $a_1 = a$, $a_2 = a + \Delta a$, $\varepsilon_1 = 1$ in Eq.(A2). Making use of linear expansion (A2) in $\Delta a/a$ we then retrieve Eq. (8) in particular case $\Delta a = \Delta r, a = r_g$.

FIGURE CAPTIONS

Fig.1 Comparison of various approximations for the photonic absorption cross –sections.
  (a) Eq.(14) –solid line, Eq. (22) –dashed line; the dotted line shows the geometrical optics limit.
  (b) Eq.(21) –solid line; the dotted, dashed and dashed-dotted lines correspond to Eq.(14) at $x = 0.001, 0.01$ and $0.1$, respectively.

Fig. 2 Schematic view of a twin-shell dielectric sphere placed in the external electric field **E**, and the coordinate system used.

FIGURE 1a

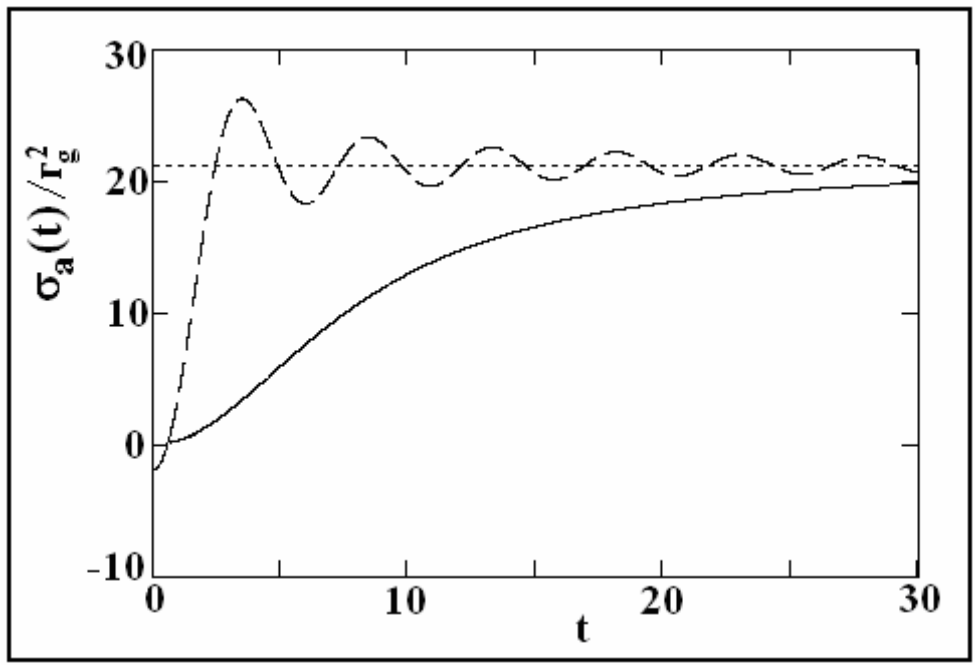

FIGURE 1b

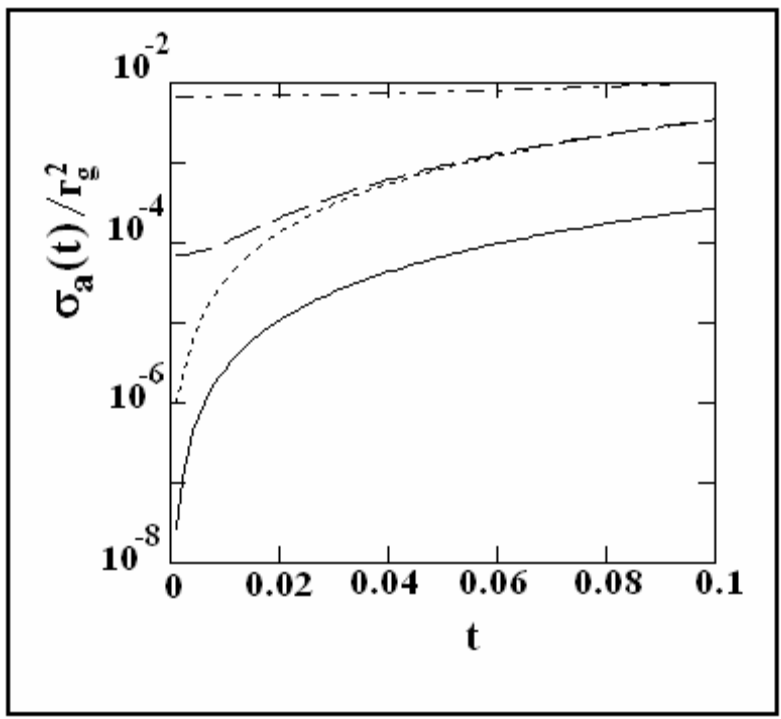

FIGURE 2

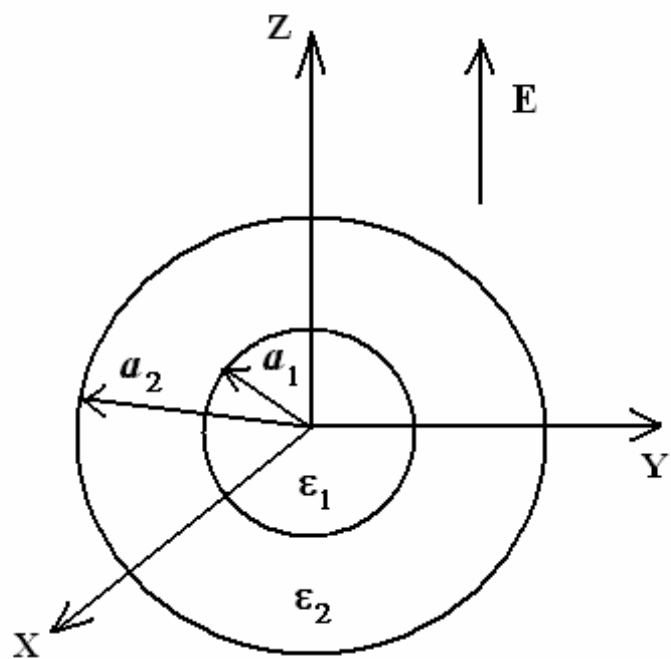